\documentclass[apj]{aastex63}

\usepackage[utf8]{inputenc}
\usepackage[T1]{fontenc}
\usepackage{ae,aecompl} 
\usepackage{graphicx}
\graphicspath{{plots/}} % Directory in which figures are stored
\usepackage{units}
\usepackage{float}
\usepackage{amsmath}
\usepackage{color}
\usepackage{amssymb}
\usepackage{latexsym}
\usepackage{wasysym}
\usepackage{psfrag}
\usepackage{ifthen}
\usepackage{longtable}
\usepackage{float}
\usepackage[utf8]{inputenc}
\usepackage{lineno}
\usepackage{multirow}
\usepackage{svg}

\newcommand{\detectionthreshold}{\unit[1/20]{$\rm yr^{-1}$}}
\newcommand{\detectionfap}{$10^{-6}$}
\begin{document}

\title[GW emission from MGFs]{Search for Long-duration Gravitational-wave Signals Associated with Magnetar Giant Flares}

\author{A. Macquet}
\affiliation{Artemis, Universit\'e C\^ote d’Azur, Observatoire de la C\^ote d’Azur, CNRS, Nice 06300, France}

\author{M.A. Bizouard}
\affiliation{Artemis, Universit\'e C\^ote d’Azur, Observatoire de la C\^ote d’Azur, CNRS, Nice 06300, France}

\author{E. Burns}
\affiliation{Department of Physics \& Astronomy, Louisiana State University, Baton Rouge, LA 70803, USA}

\author{N. Christensen}
\affiliation{Artemis, Universit\'e C\^ote d’Azur, Observatoire de la C\^ote d’Azur, CNRS, Nice 06300, France}

\author{M. Coughlin}
\affiliation{School of Physics and Astronomy, University of Minnesota, Minneapolis, Minnesota 55455, USA}

\author{Z. Wadiasingh}
\affiliation{NASA Goddard Space Flight Center, 8800 Greenbelt Road, Greenbelt, MD 20771, USA}
\affiliation{Universities Space Research Association Columbia, Maryland 21046, USA}

\author{G. Younes}
\affiliation{Department of Physics, The George Washington University, Washington, DC 20052, USA}
\affiliation{Astronomy, Physics and Statistics Institute of Sciences (APSIS), The George Washington University, Washington, DC 20052, USA}

\begin{abstract}
Magnetar giant flares are rare and highly energetic phenomena observed in the transient sky whose emission mechanisms are still not fully understood. Depending on the nature of the excited modes of the magnetar, they are also expected to emit gravitational waves, which may bring unique information about the dynamics of the excitation. A few magnetar giant flares have been proposed to be associated to short gamma-ray bursts. In this paper we revisit, with a new gravitational-wave search algorithm, the possible emission of gravitational waves from four magnetar giant flares within 5\,Mpc. While no gravitational-wave signals were observed, we discuss the future prospects of detecting signals with more sensitive gravitational-wave detectors. We in particular show that galactic magnetar giant flares that emit at least 1\% of their electromagnetic energy as gravitational waves could be detected during the planned observing run of the LIGO and Virgo detectors at design sensitivity, with even better prospects for third generation detectors.
\end{abstract}

%\maketitle

\section{Introduction}
\label{sec:introduction}

Magnetars, highly magnetized young neutron stars with surface fields often surpassing $>10^{14}$~G, are known to exhibit rare extraordinary flares characterized by micro-to-millisecond gamma-ray flashes of energies $10^{44}-10^{47}$ erg (isotropic-equivalent) followed by quasi-thermal pulsing tails at $\sim10^{44}$ erg s$^{-1}$ lasting hundreds of seconds. These magnetar giant flares are, in terms of energy, the most extreme phenomena known from isolated neutron stars. Giant flares, and other magnetar phenomenology, are thought to be powered from magnetic free energy stored internally, a reservoir up to $\sim10^{48}-10^{49}$ erg -- for reviews, see, e.g., \cite{2008A&ARv..15..225M,2015RPPh...78k6901T}.
The recent observation of GRB 200415a in the NGC 253 galaxy at 3.57 Mpc~\citep{Svinkin:2021wcp} has regenerated much discussion as to magnetar giant flares being a distinct class of short gamma-ray bursts. Its temporal and spectral uniqueness as well as the spatial coincidence with the nearby Galaxy NGC 253 results in a high likelihood of a magnetar giant flare origin. GRB 051103 and GRB 070201 are very likely also magnetar giant flares, and a recent analysis of all these events shows that GRB 070222 is probably also of this class~\citep{2021ApJ...907L..28B}. The four GRBs noted above are all within 5\,Mpc, implying a giant flare volumetric rate several orders of magnitude higher than compact object mergers~\citep{2021ApJ...907L..28B}. The large potential energy release at a relatively nearby distance might make these objects observable with gravitational waves (GWs)~\citep{2001MNRAS.327..639I,2011PhRvD..83j4014C,Quitzow_James_2017}, especially with third-generation GW detectors~\citep{Kalogera:2019bdd}.

GW searches using initial LIGO~\citep{Abbott:2007kv} data have previously been undertaken to look for events coincident with GRB 051103~\citep{abadie2012implications}, GRB 070201~\citep{abbott2008implications} and GRB 070222~\citep{aasi2014search}. No GW signals were observed, and limits were set. Likewise, limits have also been set for more common and lower energy magnetar short bursts from 6 galactic sources ~\citep{2011ApJ...734L..35A}, among which SGR 1806--20 and SGR 1900+14, are suspected to have had giant flares in the past \citep{2005Natur.434.1107P, 1999Natur.397...41H}. A limit of \unit[$2.1 \times 10^{44}$]{erg} has been set  on the GW energy emitted by SGR 1806--20 for 3 of the short bursts that occured during the Advanced LIGO's second observing run \citep{O2magnetar}.

A binary neutron star merger in M81 was excluded as the source of GRB 051103 from the lack of a GW counterpart within a \unit[[-5, +1]]{s} window around the gamma ray burst time~\citep{abadie2012implications}. Two un-modeled short duration (< \unit[1]{s}) transient {\it burst} searches were also conducted, FLARE with a \unit[[-2, +2]]{s} window~\citep{Kalmus_2007},  and X-PIPELINE with a \unit[[-120, +60]]{~s} window~\citep{Sutton:2009gi}. Assuming M81 as the source, the best upper limits for the emitted GW energy $E_{GW}$ from FLARE were \unit[$2.0 \times 10^{51}$]{erg} at \unit[100–200]{Hz} for signals of \unit[100]{ms}, and an f-mode upper limit of \unit[$1.6 \times 10^{54}$]{erg} for ringdown signals at \unit[1090]{Hz}. The X-PIPELINE limits for $E_{GW}$ were \unit[$1.2 \times 10^{52}$]{erg} at \unit[150]{Hz} and \unit[$6.0 \times 10^{54}$]{erg} at \unit[1000]{Hz}.

For GRB 070201, and assuming M31 as the host, a compact binary merger was also excluded. A search of short duration (up to \unit[0.1]{s}) GW bursts was done via a cross-correlation analysis looking for signals within a window of \unit[[-120, +60]]{s}. An upper limit on $E_{GW}$ for GW emission for bursts was set at \unit[$7.9 \times 10^{50}$]{erg} at \unit[150]{Hz}~\citep{abbott2008implications}.

GRB 070222 is assumed to come from M83, at a distance of \unit[4.6]{Mpc}. The previous analysis of LIGO data excluded a binary neutron star merger origin out to a distance of \unit[6.7]{Mpc}. For short bursts (< \unit[1]{s}), X-PIPELINE~\citep{Sutton:2009gi} was used, with a \unit[[-600, +60]]{s} window. An exclusion distance was set at \unit[8.9]{Mpc} for bursts at \unit[150]{Hz}, and \unit[3.5]{Mpc} for bursts at \unit[300]{Hz}~\citep{aasi2014search}.

Given the importance of the identification of these events as coming from magnetar giant flares, we revisit these events with a new un-modeled GW transient search pipeline, targeting long (\unit[$\gtrsim 10$]{s}) and short (\unit[$\sim 0.1-1$]{s}) signals. GRB 200415a happened after the Advanced LIGO~\citep{TheLIGOScientific:2014jea} - Advanced Virgo~\citep{TheVirgo:2014hva} observing run O3 was suspended, although there is data from GEO-HF~\citep{Dooley_2016} and KAGRA~\citep{Aso:2013} which were observing at the time. In this paper, we estimate the sensitivity of the Advanced LIGO - Advanced Virgo network to an event like GRB 200415a at O3 and design (O5) sensitivity~\citep{Abbott:2020qfu}. 
The results of a search for GW counterpart to GRB 200415a in GEO-HF and KAGRA data will be reported in a forthcoming publication of the LIGO-Virgo-KAGRA collaboration.

Section~\ref{sec:MGF} presents a description of magnetar giant flares and associated GW emission. The method by which we search for GWs associated with GRB 051103, GRB 070201, and GRB 070222 is given in Section~\ref{sec:search}. The results of the GW search is presented in Section~\ref{sec:results}. We discuss the implications of our results in Section~\ref{sec:disc}.

\section{Magnetar Giant Flares and Global Stellar Oscillations}
\label{sec:MGF} 
The pulsating tails from the three nearest magnetar giant flares are consistent with adiabatically cooling fireballs whose blackbody radii are commensurate with typical neutron star radii. As such, this phenomenology is highly indicative of disruptive activity associated with the inner magnetosphere or crust of the magnetar. Moreover, quasi-periodic oscillations (QPOs) with frequencies $\sim 20-600$ Hz (with one candidate also at 1840 Hz) imprinted on the tail light curves of two giant flares have been reported \citep{2005ApJ...628L..53I,2005ApJ...632L.111S,2006ApJ...637L.117W,2006ApJ...653..593S,2007AdSpR..40.1446W}. These QPOs are the strongest evidence of nonradial global free oscillations of the neutron stars and offer the prospect of astroseismology \citep{1998MNRAS.299.1059A,2018ASSL..457..673G} in both the electromagnetic and GW sectors.  For SGR 1806--20, the low frequency QPOs were initially thought to be long-lived for tens or hundreds of seconds. Yet, reanalyses by~\cite{Huppenkothen:2014ufa} and~\cite{2019ApJ...871...95M} have shown the frequencies damp on a short timescale $<1$ s and are consistent with $n=0$ crustal torsional or shear modes that are continually re-excited but damp by coupling to the core. This phenomenology is consistent with theory that predicts such shear modes couple more efficiently to the magnetosphere \citep{1989ApJ...343..839B, 2012ApJ...751L..41D, 2014MNRAS.443.1416G, 2021MNRAS.504.5880B} to produce detectable electromagnetic signals~\citep{2008ApJ...680.1398T}.

Nevertheless, the identification of the QPOs with known modes of neutron stars is still unclear. The low frequency $n=0$ crustal torsional or shear modes likely do not couple strongly with GWs, partly because the crust comprises only a few percent of the neutron star mass. It is not clear if other modes, potentially stronger GW emitters, are excited but are unobserved in the electromagnetic sector due to weaker coupling with the magnetosphere.
If the trigger for giant flares is internal \citep[e.g.,][]{1995MNRAS.275..255T,1998ApJ...498L..45D,2001MNRAS.327..639I,2001ApJ...561..980T,2010MNRAS.407.1926G,2015MNRAS.449.2047L,2017ApJ...841...54T} rather than magnetospheric \citep[e.g.,][]{2003MNRAS.346..540L,2007MNRAS.374..415K,2012ApJ...754L..12P}\footnote{Note that the original instability timescale estimates in \cite{2003MNRAS.346..540L} for the relativistic tearing instability mechanism were erroneous and corrected by \cite{2016MNRAS.456.3282E}. This reexamination found much shorter minimum timescales (for plausible magnetospheric parameters), commensurate with the observed rise time of giant flares.}, many possible resonant modes in the core and crust of the magnetar may be excited during giant flares depending on the nature and details of trigger mechanism. Some of these modes, such as f-modes and g-modes, may also produce GWs, though current models indicate that they will be too weak to be detected~\citep{2011MNRAS.418..659L, Zink:2011kq}. The upper limit for GW emission is ultimately derived from the magnetic free energy reservoir, and may exceed $\gtrsim 10^{48}$ erg \citep{2011PhRvD..83j4014C, 2001MNRAS.327..639I}.

Given all the uncertainties about magnetar giant flare triggering mechanisms and the local source and its possible GW emission, in the following, we conduct unmodeled GW searches with a large parameter space; we assume that the GW signal could be as short as $\sim$ \unit[.1]{s}, repeating or not over a few hundred seconds or as long-lived as $\sim$ \unit[500]{s} with a frequency range that goes up to \unit[2]{kHz}.

\section{Search methodology}
\label{sec:search}
Our sample contains $4$ GRBs for which a likely magnetar giant flare origin has been identified. For three of them, GRB 051103 \citep{2005GCN..4197....1G, hurley2010new}, GRB 070201 \citep{2007GCN..6088....1G, 2007GCN..6103....1H} and GRB 070222 \citep{2021ApJ...907L..28B, Svinkin:2016fho}, coincident data from two of the three initial LIGO detectors (H1, H2 and L1) are available from the $5^{th}$ science run (S5) from 2005 to 2007. The Virgo detector $1^{st}$ Science run (VSR1) had not started at the time of these three GRBs. GRB 200415a \citep{2020GCN.27579....1F, 2020GCN.27595....1S} has been observed after the advanced LIGO and advanced Virgo detectors had completed their $3^{rd}$ observing run (O3). Only KAGRA (K1) and the GEO-HF (G1) detectors were acquiring data at the time of the GRB. A publication by the LIGO-Virgo-KAGRA collaboration, including GRB 200415a is in preparation. We have thus not analyzed GRB 200415a, but have considered it for a prospective study to estimate the chance of detecting a GW signal when advanced GW detectors will reach their design sensitivity, expected during the $5^{th}$ observing run~\citep{Abbott:2020qfu}.

Two distinct searches are performed for each event, one targeting short-duration signals ($\sim$ \unit[0.1--1]{s}) and another that is best suited for signals with longer duration ($\sim$ \unit[10--500]{s}).
The time interval used to analyse the data for each search is referred as the \textit{on-source} window. As the time delay between gamma-ray and potential GW emission is not precisely constrained~\citep{Zink:2011kq}, we choose a large interval of \unit[512]{s} on both sides of the GRB trigger time $t_0$. We also define an \textit{off-source} window that consists in an interval of data close to (but outside of) the on-source window. These data are used to estimate the background trigger distribution and the sensitivity of the search. Both the on-source and off-source windows are split into windows of duration \unit[512]{s} that correspond to the maximal duration of the expected GW signal with $50 \%$ overlap to optimize computational efficiency. For each type of search, and each of the GRBs, whose main characteristics are given in Table~\ref{tab:events}, LIGO GW data are searched with the algorithm described below. 

\begin{table}[h]
%\footnotesize
\centering
\begin{tabular}{cccccc}
\hline
Event & Time (UTC) & Host & Distance (Mpc) & Detectors & Pair efficiency \\
\hline
GRB 051103 & 09:25:42 UTC 3 November 2005 &M$81$& $3.6$ & H2, L1 & $0.47$\\
GRB 070201  & 15:23:10 UTC 1 February 2007 &M$31$ & $0.77$& H1, H2 & $0.30$\\
GRB 070222  & 07:31:55 UTC 22 February 2007 & M$83$& $4.6$ & H1, H2 & $0.32$\\
GRB 200415a & 08:48:05 UTC 15 April 2020 & NGC $253$ &$3.3$& G1, K1 & $0.47$\footnote{Computed for the (H1, L1) pair.}\\
\hline
\end{tabular}
\caption{A summary of the magnetar giant flare samples and detectors that were observing at the time of each event. Pair efficiency is based on the quadrature sum of the detectors antenna factors and characterizes the detectors' network sensitivity to a GW signal coming from a given direction (see \cite{Thrane:2010ri}).}
\label{tab:events}
\end{table}

The GW signal is searched by cross-correlating strain data from two detectors. This method is well suited for long duration GW signals whose waveform is unknown. We use the \texttt{PySTAMPAS} pipeline~\citep{Macquet:2021}, a new python pipeline designed to search for long-duration GW signals in interferometric detectors, based on the Matlab-based \texttt{STAMP} pipeline \citep{Thrane:2010ri}. 
The pipeline parameter space searches for GW signals between $30$ and $2000$ Hz lasting from a few to $\sim$ \unit[500]{s} in the long-duration configuration, or $\sim$ \unit[1]{s} when tuned to search for shorter signals.\\

%It is based on a excess cross-power statistic and therefore requires coincident data from at least two detectors.

The data from each detector are first high pass filtered with a frequency cutoff at $22$ Hz to remove most of the low-frequency content of the detectors' strain data. A gating algorithm is applied to remove high amplitude, short duration spikes that are often present in the data~\citep{Davis:2021ecd}. A time-frequency map ($ft$-map) is built using the Fourier transform of short segments of duration \unit[1]{s}, which are Hann-windowed and overlap by \unit[50]{\%}. The frequency range of the maps is \unit[30-2000]{Hz}. The $ft$-maps are whitened by the one-sided amplitude spectral density, which is estimated by taking the median of the squared modulus of the Fourier transform over $20$ adjacent frequency bins. This estimator for the amplitude spectral density maximizes the sensitivity to monochromatic and quasi-monochromatic signals. 
Yet, GW detector data contains narrow and high amplitude spectral artefacts that have an instrumental origin (mechanical resonances, power lines among others) \cite{Davis:2021ecd}. As these lines can be easily mistaken for monochromatic GW signals, it is necessary to identify and remove the main ones. This is done using the off-source window data. For each window, frequency bins whose mean value over a map exceed a threshold of $2$ (the pixel's values follow a $\chi^2$ with 2 degrees of freedom) are listed as ``potential instrumental lines.'' If, in addition, these potential instrumental lines are persistent in more than 5\% of the total time of the off-source window, all the pixels corresponding to that frequency is set to $0$ (''notched'') for all $ft$-maps. Overall, $\sim 5\%$ of the total frequency bins are notched for each search.

A seed-based clustering algorithm is then run over the single-detector $ft$-maps to identify groups of pixels with absolute value above a given threshold~\citep{Prestegard:2016}. The parameters of the clustering algorithm are tuned according to the type of signals searched. For long-duration signals, the energy is expected to be spread over several pixels, so we set a low threshold on the individual pixels' power (incoherent energy) and dismiss clusters that contain less than $20$ pixels. The opposite approach is taken for short-duration signals. Because the energy is typically concentrated in only a few pixels, we do not apply a constraint on the number of pixel in a cluster, but we increase the threshold on pixels' incoherent energy to reduce the number of noise clusters.

The extracted clusters are cross-correlated with the other detector's pixels to compute a coherent signal-to-noise ratio (SNR).This computation takes into account the phase shift induced by the delay of arrival time of a GW signal in the two detectors, which depends on the source's position for spatially separated detectors. We consider the center of the error box given in~\cite{2005GCN..4197....1G}, \cite{2007GCN..6103....1H} and \cite{2021ApJ...907L..28B} for GRB 051103, GRB 070201 and GRB 070222 respectively.
An error over the source position induces a loss of SNR for non co-located detectors' pair, which depends on the distance between the detectors and the GW signal frequency. For GRB 051103, considering the area of the error box on the position of $120$ squared arcmins \citep{2005GCN..4197....1G}, the maximal loss of SNR would be $\sim 8 \%$ considering a GW signal at \unit[2000]{Hz}. Coherent pixels are grouped to form a \textit{trigger}. Each trigger is assigned a detection statistic $p_{\Lambda}$ based on its pixels' SNR and the incoherent energy in each detector's data~\citep{Macquet:2021}. This hierarchical approach allows to perform the analyzes much faster as cross-correlation is computed only for a small subset of pixels, without sacrificing sensitivity \citep{lonetrack}.
% BKG
To assess the significance of triggers found in the on-source window, we estimate the distribution of background triggers due to the detectors' noise. To ensure that the data do not contain any coherent GW signal, the data streams of the two detectors are shifted with respect to each other by at least \unit[256]{s}, which is much larger than the light travel time between the detectors. The data are analyzed identically to that of the on-source window, and this process is repeated for several time-shifts to simulate multiple instances of the noise. The cumulative rate of triggers provides an estimation of the false-alarm rate (FAR) that we use to determine the detection threshold on $p_{\Lambda}$ for the on-source window analysis. The FAR distributions obtained for each GRB are shown in Figure~\ref{fig:background_distributions}. The number of background triggers is different for each GRB. This is especially true for GRB 051103, which occurred at the very beginning of S5, when many sources of noise were not yet mitigated. Since loud noise triggers that populate the tail of the distributions are often due to noise fluctuations in one detector, we suppress them by requiring that the SNR ratio between the detectors be lower than $10$. The same cut is applied for the coincident analysis and efficiency estimation.
Using these results we choose a detection threshold $p_{\Lambda}$ corresponding to a FAR of $\sim$ \detectionthreshold.
Assuming a Poissonian distribution for noise events this corresponds to a false alarm probability in the on-source window of $\sim$ \detectionfap. 

%Efficiency estimation
We estimate the detection efficiency of the searches by adding simulated signals into the data. The waveforms injected are chosen to cover the parameter space constrained by the potential processes of emission described in Section~\ref{sec:MGF}. We use sinusoidal signals multiplied by a decaying exponential function (damped sine) to simulate damped, quasi-monochromatic GW emission. For long-duration searches, decay times and central frequencies of the waveforms are varied between \unit[2--10]{s} and \unit[50--500]{Hz} respectively. Short-duration searches use a damped sine with \unit[0.2]{s} decay time and a mean frequency of \unit[100]{Hz}. 
Note that the potential GW emission of MGF may be more complex than a damped sinusoid. However, since the detection method relies on a cross-correlation, the sensitivity of the search mainly depends on the frequency, duration and energy of the signal. Given the large parameter space and the lack of precisely modelled waveforms, we limit our models to damped sinusoids, which provide a good estimation of the sensitivity of the search.
%More on waveform description and injection mechanism
For each waveform, we generate 100 random starting times uniformly distributed inside the off-source window. The strain response of both detectors is computed using antenna factors and time delays that corresponds to the time and location of the targeted GRB. Since the 
orientation of the source is unknown, we draw random values uniform in the GW polarization angle and the cosine of the source's inclination angle. The simulated signals are added to the data and analyzed by the search algorithm. A signal is considered recovered if a trigger is found within the time and frequency boundaries of the simulated signal, and with a statistic $p_{\Lambda}$ higher than the detection threshold corresponding to our choice of \detectionthreshold.
This process is repeated for different signal amplitudes to estimate the detection efficiency of the search. 
%The strength of a signal is characterized by its root-sum-squared amplitude $h_{\rm rss}$, defined as
%\begin{equation}
%    h_{\rm rss} \equiv \sqrt{\int (h_+^2(t) + h_{\times}^2(t))\,dt}.
%\end{equation}

\section{Results}
\label{sec:results}
\begin{figure}[!h]
\includegraphics[scale=0.5]{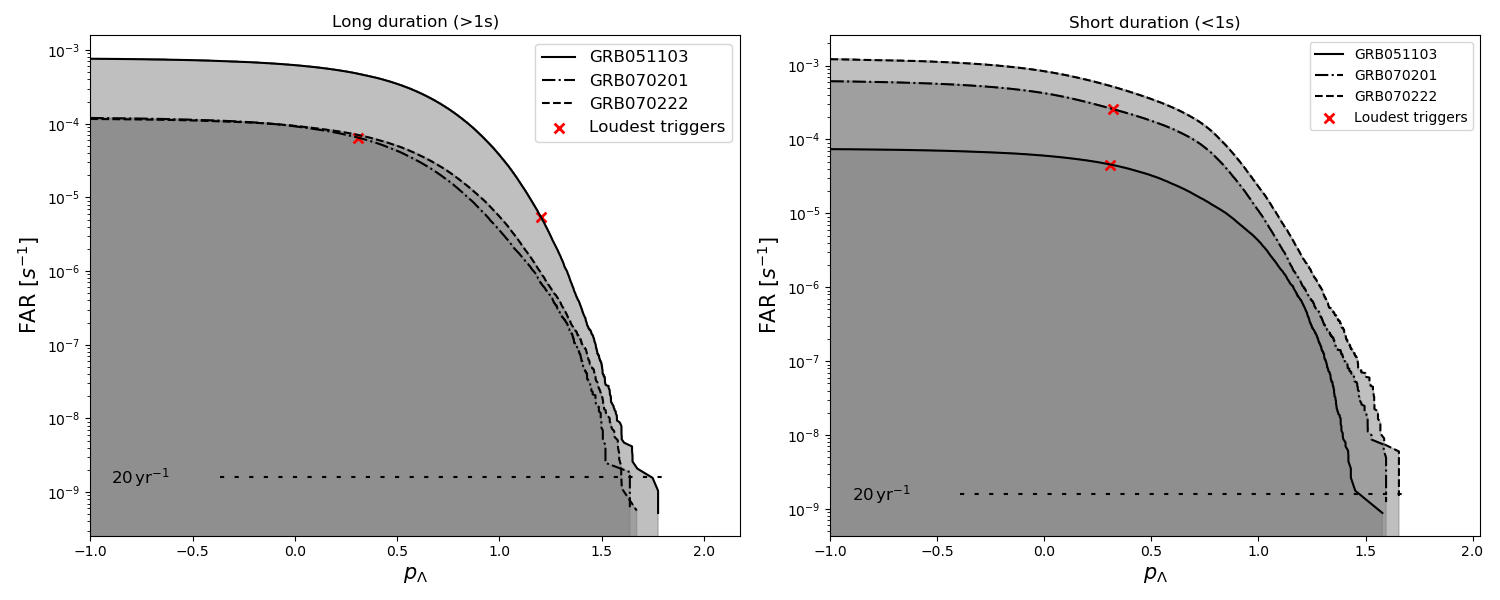}
\caption{Cumulative distribution of background noise triggers for each GRBs on top of which crosses represent the GW triggers found in the on-source windows of GRB 051103 and GRB 070201. No trigger has been found for GRB 070222. The left and right panels show the results of long-duration and short-duration searches, respectively.}
\label{fig:background_distributions}
\end{figure}

 % Zero-lag
The searches for GW signals in each GRB on-source window identified triggers; their FAR is then compared to the background estimation in Figure~\ref{fig:background_distributions}. Table \ref{table:loudest_triggers2} summarizes the characteristics of the loudest event for each GRB and each type of search. Further analysis of these triggers shows that their FARs fit well within the core of the background distribution, and their morphological properties match those of noise triggers or spectral lines. None of the triggers have a FAR lower than the detection threshold that we fixed at \detectionthreshold. We therefore report that no confident GW signal has been found in any of the GRB on-source windows in either the long and short-duration searches.

\begin{table}[!h]
\begin{center}
\begin{tabular}{cccccc}
\hline
%\multirow{2}{*}{Event} & \multicolumn{4}{c|}{Loudest trigger}\\
\cline{2-5}
 GRB & Search &$t_{start} - t_0$ (s) & Duration (s) & Frequency range (Hz) & FAP\\
 \hline
 \multirow{2}{*}{GRB 051103} & Long & $286$ & $8$ & $1640$-$1642$ & $5.5 \times 10^{-3}$\\
 & Short & $91$ & $2.5$ & 249--251 & $2.0 \times 10^{-1}$\\
 \multirow{2}{*}{GRB 070201} & Long &158 & 12 &  237--243  & $6.4 \times 10^{-2}$ \\
  & Short & $431$ & $1 $& 783--786 & $4.8 \times 10^{-1}$\\
% \multirow{2}{*}{GRB 070222} & Long &-15 & 10 & 1531--1540 & $8.8 \times 10^{-2}$ \\
%  & Short & $127$ & $1$ & 1008--1013 & $7.3 \times 10^{-1}$\\
\hline
\end{tabular}

\caption{Properties of the loudest triggers found in each GRB on-source window for long-duration and short-duration searches. $t_{start}$ refers to the starting time of the GW trigger, while $t_0$ is the GRB trigger time. The false-alarm probability (FAP) is inferred from the FAR and the duration of the on-source window.}
\label{table:loudest_triggers2}

\end{center}
\end{table}

%\subsection{Search sensitivity}

% Upper limits
We then use the results of efficiency studies to place upper limits on the emitted GW energy. 
%The strain amplitude $h_{\rm rss}$ at $50\%$ detection efficiency and a false alarm threshold of $1$ per $20$ years are reported in Figure \ref{fig:ul_hrss} for all waveforms and each GRB. 
Considering the distance to the GRBs given by the host galaxy, one can compute the isotropic GW energy radiated by a source at distance $r$ assuming a quadrupolar emission:
\begin{equation}
    E_{\rm{GW}} = r^2 \frac{c^3}{4 G} \int (\dot{h}_+^2(t) + \dot{h}_{\times}^2(t))\,dt.
\end{equation}
where $h_+(t)$ and $h_\times(t)$ are the polarisations of the GW signal. We use the simulated waveforms added to the data described in Section~\ref{sec:search} to estimate the energy corresponding to a detection efficiency of 50\% at a FAR of \detectionthreshold. These values correspond to the minimal GW energy we are able to detect in the data for each GRB. This GW energy limit is then compared to the estimation of the isotropic-equivalent electromagnetic energy of the events $E_{\rm{iso}}$ summarized in~\cite{2021ApJ...907L..28B}. All values are given in Table~\ref{tab:results}. The upper limits on $E_{\rm{GW}}$ are several orders of magnitude higher than $E_{\rm{iso}}$ and therefore do not provide meaningful constraints over the ratio of GW energy to the electromagnetic energy emitted. These results can be compared with the ones established by previous searches for short-duration GW emission around GRB 051103~\citep{Kalmus_2007, Sutton:2009gi}, GRB 070201~\citep{abbott2008implications} and GRB 070222~\citep{aasi2014search}. Overall, we report a sensitivity increase by a factor $\sim 2$ compared to upper limits set with the X-PIPELINE. For GRB 051103, FLARE reported upper limits $\sim 4$ times lower than what is obtained with \texttt{PySTAMPAS}. The discrepancy could be explained by the fact that we use a more conservative detection threshold than FLARE, which used the FAR of the loudest event in the on-source window as a detection threshold.
Furthermore, the FLARE search was sensitive to only narrow band and short duration ($<$100ms) GW signals. \texttt{PySTAMPAS} is covering a larger parameter space (in duration and bandwidth) and is sensitive to a large variety of signal morphologies.
% These results are compatible with the ones established by previous searches for short-duration GW emission around GRB 051103~\citep{Kalmus_2007, Sutton:2009gi}, GRB 070201~\citep{abbott2008implications} and GRB 070222~\citep{aasi2014search}. \mab{Actually results are better by a factor 2 except for FLARE. FAR of 1/20 yrs in this search and not the LES. Add a comment about this.}

In addition to the three GRBs analyzed using S5 data, we also performed a sensitivity study for GRB 200415a using simulated data following H1 and L1 sensitivity at the end of O3 \citep{O3bH1, O3bL1} to estimate what would have been the chance to detect GW emission from this source if the Advanced detector network was still observing in April 2020. The gain of more than two orders of magnitude observed for this GRB $E_{\rm{GW}}$ limit is mainly due to the LIGO detectors' sensitivity gain between S5 and O3. That gain would have allowed us to constrain the GW energy emitted to $E_{\rm{GW}} \lessapprox 10^{3-4} E_{\rm{iso}}$.

%\begin{figure}
%    \centering
%    \includegraphics[scale=0.5]{ul_hrss.png}
%   \caption{$h_{\rm{rss}}$ at FAD $=50\%$ and FAR = $1/10 \rm{yr}^{-1}$ for each %waveform and each search.}
%    \label{fig:ul_hrss}
%\end{figure}

{\footnotesize
\begin{table}[!h]

\begin{tabular}{c ccccc}
\hline
\multirow{2}{*}{Duration (s)} & \multirow{2}{*}{$f_0$ (Hz)} & \multicolumn{4}{c}{GW energy limits (erg)}\\
 & & GRB051103 & GRB070201 & GRB070222 & GRB200415a\\
\hline
0.2 & 100 & $1.44 \times 10^{52}$ & $3.47 \times 10^{50}$ & $1.69 \times 10^{52}$& $2.3 \times 10^{49}$\\
2 & 100 & $5.95 \times 10^{51}$ & $3.07 \times 10^{50}$ & $9.19 \times 10^{51}$& $1.43 \times 10^{49}$\\
2 & 250 &$2.56 \times 10^{52}$ & $8.83 \times 10^{50}$  & $5.56 \times 10^{52}$ &$1.21 \times 10^{50}$\\
2 & 500 & $3.32 \times 10^{53}$ & $1.25\times 10^{52}$ & $7.13\times 10^{53}$  & --\\
10 & 100 & $6.70 \times 10^{51}$ & $2.36 \times 10^{50}$  & $1.24 \times 10^{52}$ &$1.53 \times 10^{49}$\\
10 & 250 & $3.22 \times 10^{52}$ & $1.35 \times 10^{51}$  & $6.30 \times 10^{52}$ &$1.14 \times 10^{50}$\\
10 & 500 & $4.13 \times 10^{53}$ &$1.69 \times 10^{52}$ & $9.11 \times 10^{53}$ & --\\
\hline
\multicolumn{2}{c}{\multirow{2}{*}{} } & \multicolumn{4}{c}{Isotropic-equivalent EM energy $E_{\rm{iso}}$ (erg)} \\
 &  & $5.3 \times 10^{46}$ & $1.6 \times 10^{45}$ & $6.2\times 10^{45}$ & $1.3 \times 10^{46}$ \\
\hline
\end{tabular}

\caption{GW energy emitted for a source detected at 50\% efficiency for a FAP of \detectionfap. These limits are obtained considering damped sine signals whose parameters range a large portion of the parameter space. Limits for \unit[0.2]{s} emission have been obtained in the short-duration configuration of the pipeline. Values for GRB 200415a have been obtained using the last 15 days of data from O3 taken by LIGO in March 2020. We do not report values at \unit[500]{Hz} for GRB 200415a as this frequency is notched because of a mechanical resonance in H1 and L1 at this frequency during O3. The isotropic electromagnetic energy $E_{\rm{iso}}$ of each event computed by \citep{2021ApJ...907L..28B} is given for comparison.}
\label{tab:results}
\end{table}
}

\begin{figure}[!htb]
    \centering
    \includegraphics[scale=0.5]{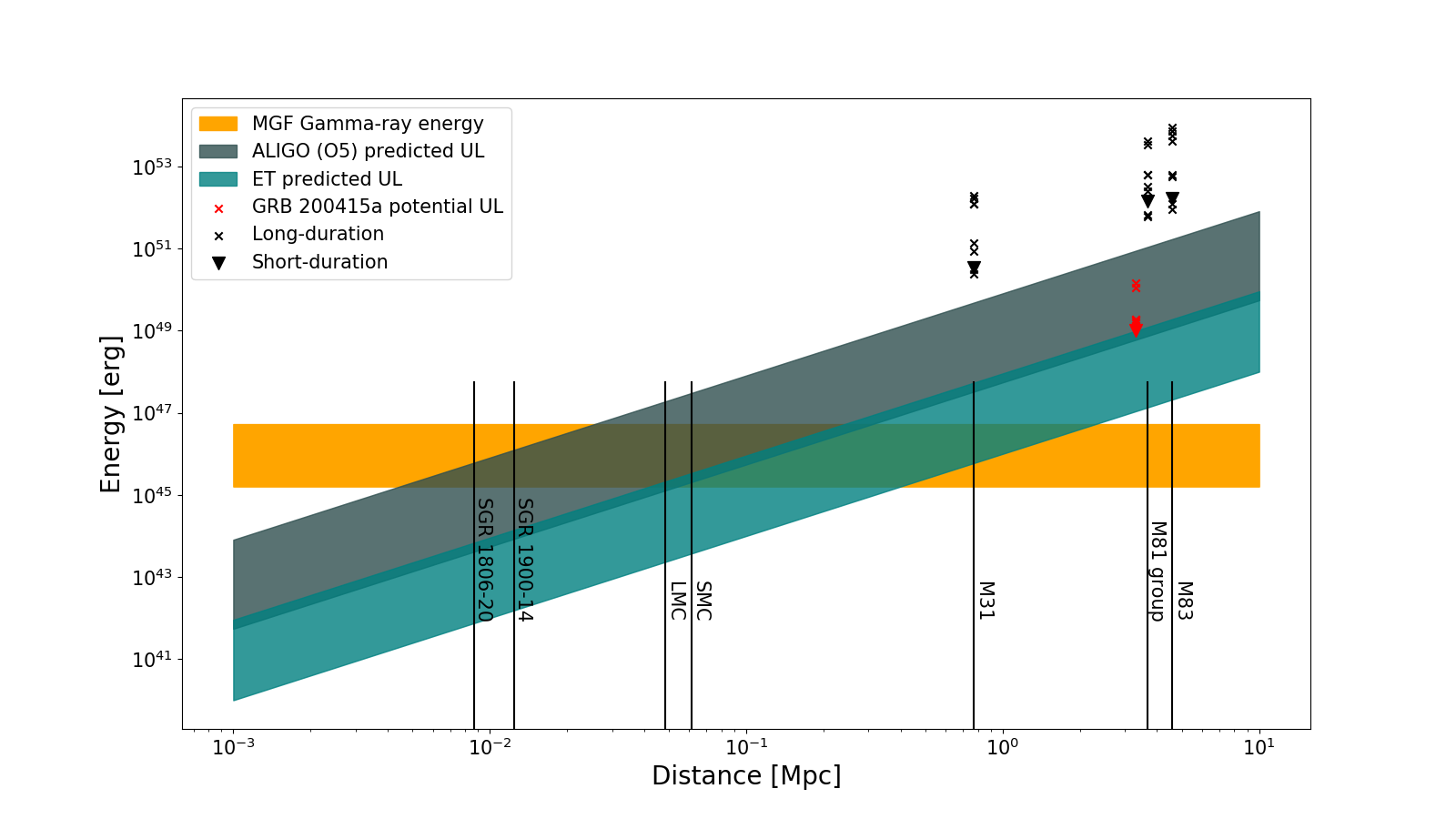}
    \caption{Upper limits (UL) on the GW energy emitted by GRB051103 (M81), GRB070222 (M83) and GRB070201 (M31) during S5 are represented with cross markers for long duration emission, and triangular markers for short duration emission. Each point corresponds to a different duration/frequency damped sine simulated waveform added to the data. For GRB200415a we report with the cross markers the limits that could have been set up for this event with LIGO if it was observing at the time. The dark grey band and blue band represent the limit on $E_{\rm{GW}}$ as function of the distance to the source (averaging over all possible sky positions) that will be achievable with Advanced LIGO data at design sensitivity (O5) and Einstein Telescope respectively. The orange band represents the minimal/maximal $E_{\rm{iso}}$ estimated for the three GRBs analyzed in this study. Distances of nearest galaxies and Soft Gamma Repeaters SGR 1806--20 and SGR1900--14 are shown for convenience.}
    \label{fig:UL}
\end{figure}

Finally, to estimate the detection sensitivity that will be achievable in the future, we used simulated data following Advanced LIGO's design sensitivity (expected during the O5 run), and the sensitivity of a proposed 3rd generation GW detector Einstein Telescope~\citep{ET}, which is expected to be in a network with the US-led Cosmic Explorer~\citep{reitze2019cosmic}. As the detectors' network sensitivity depends on the direction of the GW source, waveforms have been injected at random positions to simulate a generic magnetar giant flare source. In this study, we used the Einstein Telescope, treated as two co-located interferometric detectors whose arms form a $60^{\circ}$ angle and arbitrarily located in Italy, but similar results would have been obtained with two Cosmic Explorer-like detectors in the USA. We summarize all these results in Figure~\ref{fig:UL} and compare the inferred upper limits on the GW energy detectable to the estimated $E_{\rm{iso}}$ of the candidate magnetar giant flares as a function of the source's distance. Upper limits on the GW energy decrease by a factor $\sim 2$ between O3 and O5, which is compatible with the detector's expected sensitivity gain \citep{Davis:2021ecd}. 
We show that it would be possible to constrain the GW energy emitted up to a fraction of $E_{\rm{iso}}$ for magnetar giant flares in the Milky Way and in the Magellanic Clouds, in both scenarios of long-duration and short-duration GW emission with Advanced LIGO at design sensitivity. Third generation detectors such as the Einstein Telescope or Cosmic Explorer should provide a factor $\sim 100$ increase in sensitivity to GW energy, making even lesser energetic flares from galactic magnetars detectable.

%\subsection{Future sensitivity}
% Future sensitivity

\hfill
\section{Discussion}
\label{sec:disc}

The recent observation of GRB 200415a, suggesting that magnetar giant flare may be a distinct class of short gamma ray bursts, with a substantially higher volumetric rate than compact object mergers \citep{2021ApJ...907L..28B}, has motivated the re-examination of gravitational-wave data around three other likely magnetar giant flare events in nearby galaxies. We used a new GW search pipeline \texttt{PySTAMPAS}~\citep{Macquet:2021}, that allows us to cover many possible GW emission mechanisms. Given the large distances to these sources and the typical electromagnetic energy emitted, the prospects of detecting such signals from outside the Milky Way were low. Yet substantial and imminent improvements in GW detector sensitivity will enable probing a regime where the GW energy is comparable or less than the typical magnetar giant flare electromagnetic energy for the first time for extragalactic events. Moreover, future Galactic giant flares may lead to the first detection under the assumption that the GW signal carries >~0.1\% of the EM energy released, with limits down to $\gtrsim 10^{-6}$ for third generation detectors, which can probe GWs for even the weakly-GW-emitting low frequency torsional modes. For the nearest Galactic magnetars at $\sim$ 2 to 4 kpc, third generation detectors could even begin probing a regime of more common recurrent magnetar short bursts and intermediate flares, which have electromagnetic energies $\lesssim 10^{43}$ erg \citep{2008ApJ...685.1114I}, especially during burst forests when repeated excitations are plausible \citep{2020ApJ...904L..21Y}. Our new pipeline readies us for such an era. 

As mentioned in~\S\ref{sec:MGF}, the essential trigger of magnetar giant flares is unknown, although many proposals exist \citep[e.g.,][]{1980Natur.287..122R,1995MNRAS.275..255T,1998ApJ...498L..45D,2001MNRAS.327..639I,2001ApJ...561..980T,2003MNRAS.346..540L,2007MNRAS.374..415K,2010MNRAS.407.1926G,2012ApJ...754L..12P,2015MNRAS.449.2047L,2016MNRAS.456.3282E,2017ApJ...841...54T}. The high (volumetric) rate of giant flares \citep{2021ApJ...907L..28B}, which exceeds the core collapse supernova rate, implies giant flares may reoccur many times during a magnetar's active lifespan.  This disfavors giant flares mechanisms with finality, such as phase transitions \citep{1980Natur.287..122R},  redistribution of stellar structure and moment of inertia  \citep{2001MNRAS.327..639I} or catastrophic internal magnetic field rearrangements such as the magnetohydrodynamic interchange instability \citep{1995MNRAS.275..255T}. These dramatic internal mechanisms may produce GWs with considerable power, possibly comparable to a few percent of gravitational binding energy of $\sim 10^{53}$ erg. Our limits on the M31 giant flare GRB 070201 encroach on such an energy scale, and future limits on nearby giant flares will definitively rule out such energetic scenarios. 

Note that the location and nature of the physical trigger mechanism, for instance whether it is magnetospheric or internal to the neutron star, can influence the character of the global free oscillations and what modes may be excited in a rich but complicated manner that depends on the equation of state and configuration of magnetic fields in the core and crust of the magnetar \citep{2007PhRvD..75h4038P,2009MNRAS.396.1441C,2011MNRAS.414.3014C,2012MNRAS.423..811C,2011PhRvD..83j4014C,Zink:2011kq,2011MNRAS.410L..37G,2012MNRAS.421.2054G}. This can strongly influence the relative apportionment between energy emitted in the gravitational and electromagnetic sectors, particularly if f-modes are excited more efficiently over torsional or shear modes that couple more readily to the magnetosphere for electromagnetic emission \citep{1989ApJ...343..839B,2008ApJ...680.1398T,2014MNRAS.441.2676L,2021MNRAS.504.5880B}. Yet, significant theoretical uncertainty also exists on the EM emission locale, radiative processes, photon transport, beaming and outflows in giant flares \citep[e.g.,][]{2016MNRAS.461..877V} which can muddle firm inferences on the EM energetics. 
Thus characterization or constraints on GWs from magnetar activity is essential to understanding the energetics and nature of the unknown trigger(s) and how strongly they couple to the interior of the neutron star. This has implications beyond giant flares, as more common short bursts may also share the same physical trigger and similar QPOs have been reported in those \citep{2014ApJ...795..114H,2014ApJ...787..128H}. Likewise, the inferred power-law energy distribution of giant flares reported by \cite{2021ApJ...907L..28B} is consistent swith the recurrent short burst distribution observed in Galactic magnetars, suggesting a continuum of bursts energies and a similar trigger for both lower-energy recurrent short bursts and giant flares. Recent unprecedented results from NICER of SGR 1830--0645 (Younes et al. submitted) also reveal a strong phase dependence of short bursts aligned with the surface thermal emission pulse profile; this would point to a low-altitude trigger, associated with the crust, and might disfavor high-altitude equatorial magnetospheric trigger models. A low-altitude or crustal trigger would perhaps improve prospects for third generation GW detectors, particularly for limits approaching the level of $\sim10^{-6}$ of the EM power expected from excited f-modes \citep{2011MNRAS.418..659L}.
However, given the uncertainties, a non-detection may not be enough to rule out most of the models describing the coupling between the crust and the core.
Nevertheless, after the recent association of a Fast Radio Burst with a similar short burst event \citep{2020Natur.587...59B,2020ApJ...898L..29M,2020arXiv200511071L}, and proposals that magnetar oscillations may underlie some fast radio bursts \citep{2019ApJ...879....4W,Wadiasingh2020,2019MNRAS.488.5887S,2020ApJ...903L..38W}, GW studies offer a potentially unique view on the trigger of magnetar bursts. 
This also motivates the development of new detection algorithms that more specifically targets repeating signals associated with QPOs.

Finally, we note that the volumetric high rate of magnetar giant flares inferred by \cite{2021ApJ...907L..28B} potentially allows for a non-negligible contribution to the stochastic GW background~\citep{Christensen_2018} by mature magnetars, likely pertinent for third generation detectors such as the Einstein Telescope~\citep{ET} and Cosmic Explorer~\citep{reitze2019cosmic}. This will be investigated in a future study.

\bigskip\noindent\textit{Acknowledgments} ---
N.C. is supported by NSF grant PHY-1806990. M.C. is supported by NSF grant PHY-2010970. Z.W. is supported by the NASA postdoctoral program. This research has made use of data, software and/or web tools obtained from the Gravitational Wave Open Science Center (https://www.gw-openscience.org/), a service of LIGO Laboratory, the LIGO Scientific Collaboration and the Virgo Collaboration. LIGO Laboratory and Advanced LIGO are funded by the United States National Science Foundation (NSF) as well as the Science and Technology Facilities Council (STFC) of the United Kingdom, the Max-Planck-Society (MPS), and the State of Niedersachsen/Germany for support of the construction of Advanced LIGO and construction and operation of the GEO600 detector. Additional support for Advanced LIGO was provided by the Australian Research Council. Virgo is funded, through the European Gravitational Observatory (EGO), by the French Centre National de Recherche Scientifique (CNRS), the Italian Istituto Nazionale di Fisica Nucleare (INFN) and the Dutch Nikhef, with contributions by institutions from Belgium, Germany, Greece, Hungary, Ireland, Japan, Monaco, Poland, Portugal, Spain~\citep{Rich_Abbott_2021}. The authors are grateful for computational resources provided by the LIGO Laboratory and supported by National Science Foundation Grants PHY-0757058 and PHY-0823459.

\bibliography{bibliography}

\end{document}